\documentclass[letterpaper,conference,twocolumn]{IEEEtran}
\usepackage{ifpdf}
\usepackage{color}
\usepackage[latin1]{inputenc}
\usepackage{cite}
\usepackage{mathtools}
\usepackage{amssymb}
\usepackage{mathtools}
\usepackage{algorithm}
\usepackage{algpseudocode}

\usepackage{subfigure}

\ifpdf
\usepackage{graphicx} \DeclareGraphicsExtensions{.png} \graphicspath{{Figures}}
\else
\usepackage{graphicx} \DeclareGraphicsExtensions{.pdf} \graphicspath{{Figures}}
\usepackage{graphicx} \DeclareGraphicsExtensions{.eps} \graphicspath{{Figures}}
\fi

\usepackage{tikz}
\usetikzlibrary{decorations.pathmorphing}
\usetikzlibrary{fit}
\usetikzlibrary{backgrounds}
\usetikzlibrary{shapes}
\usetikzlibrary{positioning}
\usetikzlibrary{shadows}
\usetikzlibrary{decorations.pathmorphing}
\usetikzlibrary{decorations.shapes}
\usepackage{multirow}
\usepackage{dcolumn}
\usepackage{pgfplots}
\usepackage{flushend}
\pgfplotsset{compat=1.3}

\begin{document}

\title{Packet Travel Times in Wireless Relay Chains under\\ Spatially and Temporally Dependent Interference}
\author{Alessandro Crismani$^1$, Udo Schilcher$^1$, Stavros Toumpis$^2$, G\"{u}nther Brandner$^1$ and Christian Bettstetter$^{1,3}$ \\
        $^1$ University of Klagenfurt, Networked and Embedded Systems, Klagenfurt, Austria\\
        $^2$ Athens University of Economics and Business, Department of Informatics, Athens, Greece \\
        $^3$ Lakeside Labs GmbH, Klagenfurt, Austria\\
        email: \texttt{alessandro.crismani@aau.at}
}

\markboth{}{}
\thispagestyle{empty}

\newcounter{mytempeqncnt}

\newcommand{\etal}{\textit{et al.\;}}
\newcommand{\expect}{{\rm E}}
\newcommand{\var}{{\rm Var}}
\newcommand{\cov}{{\rm Cov}}
\newcommand{\prob}{{\rm P}}
\newcommand{\indic}{{\rm\bf 1}}
\newcommand{\dd}{\:\:{\rm d}}
\newcommand{\eul}{\mathrm{e}}
\newcommand{\ppp}{\Phi}
\newcommand{\pppt}{\Phi_t}
\newcommand{\txp}{p}
\newcommand{\dens}{\lambda}
\newcommand{\thr}{\theta}
\newcommand{\thrL}[1]{\theta_{{#1}}}
\newcommand{\covbound}{C}

\newcommand{\source}{S}
\newcommand{\relay}[1]{R_{{#1}}}
\newcommand{\destination}{D}
\newcommand{\tx}[1]{t_{{#1}}}
\newcommand{\rx}[1]{r_{{#1}}}
\newcommand{\link}[1]{\mathcal{L}_{{#1}}}
\newcommand{\hop}{L}
\newcommand{\ch}[1]{h({#1})}
\newcommand{\chu}[1]{h_u({#1})}
\newcommand{\loss}[2]{g({#1},{#2})}
\newcommand{\rxpow}[1]{P_{#1}}
\newcommand{\intL}[1]{I_{#1}}
\newcommand{\sir}[1]{\rho_{#1}}
\newcommand{\ualoha}{\mathrm{1}[u \in \pppt]}

\newcommand{\EvSucc}[1]{\mathcal{S}_{#1}}
\newcommand{\Psucc}[1]{\Omega_{#1}}
\newcommand{\T}{T}
\newcommand{\TL}[1]{T_{#1}}
\newcommand{\tL}[1]{t_{#1}}
\newcommand{\rtx}[2]{\mathcal{S}_{#1}^{#2}}
\newcommand{\rtxC}[2]{\overline{\mathcal{S}_{#1}^{#2}}}
\newcommand{\speed}{V_p}

\sloppy

\maketitle

\thispagestyle{empty}

\begin{abstract}
We investigate the statistics of the number of time slots $\T$ that it takes a packet to travel through a chain of wireless relays. Derivations are performed assuming an interference model for which interference possesses spatiotemporal dependency properties.
When using this model, results are harder to arrive at analytically, but they are more realistic than the ones obtained in many related works that are based on independent interference models.

First, we present a method for calculating the distribution of $\T$. As the required computations are extensive, we also obtain simple expressions for the expected value $\expect [\T]$ and variance $\var [\T]$. Finally, we calculate the asymptotic limit of the average speed of the packet. Our numerical results show that spatiotemporal dependence has a significant impact on the statistics of the travel time $T$. In particular, we show that, with respect to the independent interference case, $\expect [\T]$ and $\var[\T]$ increase, whereas the packet speed decreases.
\end{abstract}

\section{Introduction}
This paper studies the transport of a packet through a multi-hop chain of nodes, under interference that exhibits spatial and temporal dependency. In particular, we derive the statistics of the number of time slots $\T$ required to travel across the chain, in a scenario where multiple retransmissions due to failures on each link may delay the packet reception. We consider interferers located on the plane according to a Poisson point process (PPP), and we assume that their positions do not change over time. This modeling assumption leads to temporally and spatially dependent interference, since the interference powers at different locations and times are influenced by the same set of nodes.

Our contributions are as follows:
\begin{itemize}
    \item We provide an expression for the probability mass function (PMF) of $\T$.
    \item We derive and study the first two moments of $\T$; we compare the results for the case of having temporally and spatially dependent interference with the somewhat less realistic scenario where the interference levels at different transmissions are independent.
    \item We define the average speed of a packet traveling the relay chain, and we derive its asymptotic limit when the number of relays grows to infinity.
    \item Ultimately, we show that the model adopted for the interference significantly changes the estimated network performance; in particular, relying on the assumption that interference is independent at different times and locations leads to an underestimation of the values of $\expect [\T]$ and, especially, $\var [\T]$, as well as an overestimation of the asymptotic limit of the average speed.
\end{itemize}

The paper is organized as follows. Section~\ref{sec:related} briefly summarizes the literature that is related to this work. Section~\ref{sec:model} details the system under study and presents our modeling assumptions. The PMF of $\T$ is derived in Section~\ref{sec:pmf}. Since computing the PMF is infeasible for large networks, in Sections~\ref{sec:av-time} and~\ref{sec:var-time}, we derive simple expressions for the expected value and variance of $\T$. Section~\ref{sec:av-var-results} compares these values for temporally and spatially dependent or independent interference. Section~\ref{sec:speed} discusses the average speed of a packet traveling through the relay chain. Finally, Section~\ref{sec:conclusions} concludes.

\section{Related Work}
\label{sec:related}
Relay chains have attracted significant research interest. The authors of~\cite{xie2004relay} consider a node chain and derive the achievable coding rate under the assumption of a discrete memoryless channel. The authors of~\cite{farhadi2009multihop} derive the ergodic capacity for such chains. The diversity achieved in a multi-hop relay network is discussed in~\cite{boyer2004mutlihop}. These papers, however, do not consider the influence of interference on system performance.

The influence of co-channel interference generated by nodes located according to a PPP is studied for single-hop networks in~\cite{haenggi09:interference,haenggi09:outage,ganti09:interf-correl,schilcher12:interfcor,tanbourgi2013coop,schilcher2013coop,schilcher2013temporal,haenggi2014divpol}. Multi-hop networks are considered in~\cite{bacelli2006aloha}, where the authors derive the average progress of a packet toward its destination. Such PPP networks, where transmissions interfere with each other, have also been studied in~\cite{haenggi2013local} and~\cite{haenggi2013moblocal}, where the authors analyze the local delay of interference-limited mobile scenarios. The end-to-end delay of a Poisson point process network is discussed in~\cite{haenggi2013delay}.

\section{Modeling Assumptions}
\label{sec:model}
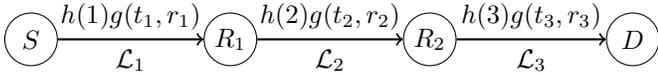
\begin{figure}[t]
    \centering
    \begin{tikzpicture}[auto]
        \pgfplotsset{tick style={draw=none}}
            \node[black, draw, circle, minimum size=0.7cm, inner sep=1pt] at  (0,0) (source) {$\source$};
            \node[black, draw, circle, minimum size=0.7cm, inner sep=1pt] at (8,0) (destination) {$\destination$};
            \node[draw, circle, minimum size=0.7cm, inner sep=1pt] at (2.65,0) (r1) {$\relay{1}$};
            \node[draw, circle, minimum size=0.7cm, inner sep=1pt] at (5.3,0) (r2) {$\relay{2}$};
            \draw[->, thick] (source) -- node[midway] {$\ch{1} \loss{\tx{1}}{\rx{1}}$} node[midway, anchor=north] {$\link{1}$} (r1);
            \draw[->, thick] (r1) -- node[midway] {$\ch{2} \loss{\tx{2}}{\rx{2}}$} node[midway, anchor=north] {$\link{2}$} (r2);
            \draw[->, thick] (r2) -- node[midway] {$\ch{3} \loss{\tx{3}}{\rx{3}}$} node[midway, anchor=north] {$\link{3}$} (destination);
    \end{tikzpicture}
    \caption{Example network scenario with two relays.}
    \label{fig:network}
\end{figure}

We consider a wireless scenario where a packet sent from a source $\source$ toward a destination $\destination$ travels through a chain of $(N-1)$ relays $\left\{ \relay{n} \right\}_{n=1}^{N-1}$. The network is composed of $N$ links $\{ \link{n} \}_{n=1} ^ {N}$ that connect $\source$ and $\destination$. We let $\source, \destination, \relay{n}$ denote both the nodes and their coordinates on the plane. The transmitter and the receiver that are connected by $\link{n}$ are called $\tx{n}$ and $\rx{n}$, respectively. This notation implies that $\source = \tx{1}$ and $\destination = \rx{N}$. Again, $\tx{n}$ and $\rx{n}$ denote both the nodes and their positions. We assume that relays have known, and not random, positions. An example network with two relays is shown in Fig.~\ref{fig:network}.

Time is slotted, and the transmission sequence develops as follows. In the first time slot, $\source$ transmits the packet and $\relay{1}$ tries to receive it. If the decoding at $\relay{1}$ fails, $\source$ retransmits the packet in the following slot. Retransmissions continue until $\relay{1}$ successfully receives the packet. Then, $\relay{1}$ transmits the packet to $\relay{2}$, following the same procedure. The transmission sequence continues until the packet forwarded by $\relay{N-1}$ is correctly decoded at $\destination$.

Transmissions are subject to co-channel interference. Interferers are distributed on the plane according to a PPP $\ppp$ of intensity $\dens$~\cite{stoyan95}. Let $\ppp$ also denote the locations of the interferers. Each interferer accesses the channel at each time slot using slotted ALOHA, i.e., it transmits at each time slot with probability $\txp$. We denote the set of active interferers in a time slot with $\pppt$.

Two scenarios are compared: the dependent interference one, where the positions of the interferers remain the same for the whole transmission sequence, and the independent interference one, where the positions change at every time slot according to a new realization of the PPP. As opposed to assuming independent interference, the dependent interference model is a somewhat more realistic model for the temporal and spatial dependency of interference under the given assumptions.

The channel connecting $\tx{n}$ and $\rx{n}$ is modeled using a distance-dependent path loss combined with Rayleigh fading. In particular, the power that is received at the receiver from the transmitter is
\begin{equation}
    \rxpow{n} = \ch{n} \loss{\tx{n}}{\rx{n}},
    \label{eqn:rx_power}
\end{equation}
where $\ch{n}$ models Rayleigh fading and is an exponentially distributed random variable having both mean and variance equal to unity. The parameter $\loss{\tx{n}}{\rx{n}}$ represents path loss. In our model, the path loss parameter captures the variance of the received power due to the distance between two stations, while the Rayleigh fading accounts for micro-mobility. We also denote with $\chu{n}$ the fading coefficient of the link connecting the interferer $u \in \ppp$ to $\rx{n}$ and with $\loss{u}{\rx{n}}$ the corresponding path loss value. We assume that the values of $\ch{n}$ and of $\chu{n}$ at different time slots are independent. For simplicity, we do not explicitly mention the time indices of fading values. Furthermore, we assume that the fading values of two links connecting two different node pairs are independent, even when they refer to the same time slot and when the two pairs share a single common node.

Our modeling assumptions can be classified according to the $(i,j,k)$ notation introduced in~\cite{schilcher12:interfcor} for the nodes' positions, the channel evolution and the traffic behavior, respectively. In particular, our dependent interference model has coordinates $(2,1,1)$, since we assume static but unknown positions of interferers ($i=2$), a channel that changes randomly every slot ($j=1$), and ALOHA $(k=1)$. The independent interference model has coordinates $(1,1,1)$, since the positions of interferers change randomly every slot $(i=1)$, and the same channel and traffic models are used.

We consider an interference limited scenario, hence we neglect the effects of noise. The signal-to-interference ratio (SIR) at $\rx{n}$ at the considered time slot can be expressed as
\begin{equation}
    \sir{n} = \frac{\rxpow{n}}{\intL{n}} = \frac{\ch{n} \loss{\tx{n}}{\rx{n}}}{\sum_{u \in \ppp} \chu{n} \loss{u}{\rx{n}} \ualoha},
    \label{eqn:sir}
\end{equation}
where $\intL{n} = \sum_{u \in \ppp} \chu{n} \loss{u}{\rx{n}} \ualoha$ is the total interference power received at $\rx{n}$.

Finally, we assume that the transmission technology adopted is such that $\rx{n}$ correctly decodes the received packet if and only if $\sir{n}$ is higher than a threshold $\thr$. The ratio between the success threshold and the path loss between $\tx{n}$ and $\rx{n}$ is
\begin{equation}
    \thrL{n} = \frac{\thr}{\loss{\tx{n}}{\rx{n}}}.
    \label{eqn:thresh-link}
\end{equation}

\section{PMF of Packet Travel Time}
\label{sec:pmf}
\begin{figure*}[!t]
    \setcounter{mytempeqncnt}{\value{equation}}
    \setcounter{equation}{5}
    \begin{align}
        &\prob \left[ \TL{1} = \tL{1}, \dots, \TL{N} = \tL{N} \right] = \expect_\ppp \left[ \prob \left[ \TL{1} = \tL{1}, \ldots, \TL{N} = \tL{N} | \ppp \right] \right] = \expect_\ppp \left[ \prob \left[ \TL{1} = \tL{1} | \ppp \right] \dots \prob \left[ \TL{N} = \tL{N} | \ppp \right] \right] \notag \\
        &= \sum_{i_1=0}^{\tL{1}-1} \dots \sum_{i_{N}=0}^{\tL{N}-1} (-1)^{i_1 + \dots + i_{N}} \binom{\tL{1} \! - \! 1}{i_1} \dots \binom{\tL{N}\! - \! 1}{i_N}\expect_{\ppp} \left[ \prod_{u \in \ppp} \! \left( \frac{\txp}{1 + \thrL{1} \loss{u}{\rx{1}}} + (1 - \txp) \! \right) \dots \left( \frac{\txp}{1 + \thrL{N} \loss{u}{\rx{N}}} + (1 - \txp) \! \right) \cdot \right. \notag \\
        & \left. \left( \frac{\txp}{1 + \thrL{1} \loss{u}{\rx{1}}} + (1 - \txp) \right) ^ {\tL{1}-1-i_1} \dots \left( \frac{\txp}{1 + \thrL{N} \loss{u}{\rx{N}}} + (1 - \txp) \right) ^ {\tL{N}-1-i_{N}} \right]. \notag \\
        &\overset{(a)}{=} \sum_{i_1=0}^{\tL{1}-1} \dots \sum_{i_{N}=0}^{\tL{N}-1} (-1)^{i_1 + \dots + i_{N}} \binom{\tL{1}-1}{i_1} \dots \binom{\tL{N}-1}{i_N} \exp \left(-\dens \int_{\mathbb{R}^2} \left[ 1 - \left( \frac{\txp}{1 + \thrL{1} \loss{x}{\rx{1}}} + (1 - \txp) \right) \cdot \right. \right. \notag \\
        & \dots \left( \frac{\txp}{1 + \thrL{N} \loss{x}{\rx{N}}} + (1 - \txp) \right) \left. \left. \left( \frac{\txp}{1 + \thrL{1} \loss{x}{\rx{1}}} + (1 - \txp) \right) ^ {\tL{1}-1-i_1} \!\!\!\!\! \dots \left( \frac{\txp}{1 + \thrL{N} \loss{x}{\rx{N}}} + (1 - \txp) \right) ^ {\tL{N}-1-i_{N}} \right] \dd x \right).
        \label{eqn:pmf-part}
    \end{align}
    \setcounter{equation}{\value{mytempeqncnt}}
    \hrule
\end{figure*}
The random variable representing the number of time slots required to traverse the chain is called $\T$. We can express $\T$ as
\begin{equation}
    \T = \sum_{n=1}^{N} \TL{n},
    \label{eqn:succ-time}
\end{equation}
where $\TL{n}$ is the number of time slots required to traverse link $\link{n}$. We are interested in calculating the PMF of $\T$. We have
\begin{equation}
    \prob \left[ \T = t \right] = \sum_{ \tL{1} + \dots + \tL{N} = t } \prob \left[ \TL{1} = \tL{1}, \ldots, \TL{N} = \tL{N} \right],
    \label{eqn:pmf}
\end{equation}
where the positive integer $\tL{n} \geq 1$ represents the number of transmissions on link $\link{n}$.

Let us consider a single link $\link{n}$, and denote the event that its $i$-th transmission is successful as $\rtx{n}{i}$. In the case of dependent interference, the locations $\ppp$ of interferers remain the same for all transmissions. Since these locations are random, the path loss values $\loss{u}{\rx{n}}$ are random variables. Moreover, the path loss values at different transmissions are the same, since the positions of interferers do not change. Hence, the success events of two transmissions on link $\link{n}$ are dependent, since they are a function of the path loss values. However, if we consider a given realization of $\ppp$, that is if we fix the positions of the interferers, the path loss values are not random. Hence, given a realization of the PPP, two transmissions $i$ and $j$ at different time slots are independent, since the only remaining random contributions are from the fading values $\ch{n}$ and $\chu{n}, \forall u \in \ppp$, which we assumed to be independent at different time slots, and from the channel access behavior of interferers, which is independent between two time slots since they adopt slotted ALOHA. Hence, the probability that $\link{n}$ is successful after exactly $\tL{n}$ transmissions is
\addtocounter{equation}{1}
\begin{align}
    &\prob \left[ \TL{n} = \tL{n} \right] = \expect_{\ppp} \left[ \prob \left[ \rtxC{n}{1}, \dots, \rtxC{n}{\tL{n}-1}, \rtx{n}{\tL{n}} | \ppp \right] \right] \notag \\
    &= \expect_{\ppp} \left[ \prob \left[ \rtxC{n}{1} | \ppp \right] \dots \prob \left[ \rtxC{n}{\tL{n}-1} | \ppp \right] \prob \left[ \rtx{n}{\tL{n}} | \ppp \right] \right],
    \label{eqn:succ-rtx}
\end{align}
where $\rtxC{n}{i}$ means that the $i$-th transmission was unsuccessful.

The probability $\prob \left[ \rtx{n}{i} | \ppp \right]$ that the $i$-th transmission is successful, once conditioned on the locations of interferers, is
\begin{align}
    &\prob \left[ \rtx{n}{i} | \ppp \right] \! =  \! \prob \! \left[ \ch{n} \loss{\tx{n}}{\rx{n}} \! > \! \thr \sum_{u \in \ppp} \chu{n} \loss{u}{\rx{n}} \ualoha \right] \notag \\
    &\overset{(a)}{=} \expect_{\chu{n}} \left[ \exp \left(- \thrL{n} \sum_{u \in \ppp} \chu{n} \loss{u}{\rx{n}} \ualoha \right) \right] \notag \\
    &= \expect_{\chu{n}} \left[\prod_{u \in \ppp} \exp  \left(- \thrL{n} \chu{n} \loss{u}{\rx{n}} \ualoha \right) \right] \notag \\
    &\overset{(b)}{=} \prod_{u \in \ppp} \expect_{\chu{n}} \left[ \exp \left(- \thrL{n} \chu{n} \loss{u}{\rx{n}} \ualoha \right) \right] \notag \\
    &\overset{(c)}{=} \prod_{u \in \ppp} \left( \frac{\txp}{1 + \thrL{n} \loss{u}{\rx{n}}} + (1 - \txp) \right),
    \label{eqn:succ-prob-slot}
\end{align}
where the fading values $\ch{n}$ and $\chu{n}$ refer to the time slot on which the $i$-th transmission over $\link{n}$ takes place, and $\mathrm{1}[(\cdot)]$ is the indicator function. In~\eqref{eqn:succ-prob-slot}, $(a)$ is obtained by conditioning on the fading values of interferers and on the slotted ALOHA access and by using the complementary cumulative distribution function of $\ch{n}$, $(b)$ follows from having independent fading values from interferers and from the fact that nodes transmit independently of each other, under the slotted ALOHA scheme and, finally, in $(c)$ we have calculated the expected value with respect to fading and slotted ALOHA. We also note that~\eqref{eqn:succ-prob-slot} does not depend on the particular slot considered and hence does not depend on the transmission index $i$.

By substituting~\eqref{eqn:succ-prob-slot} into~\eqref{eqn:succ-rtx} we have
\begin{align}
    &\prob \left[ \TL{n} = \tL{n} \right] = \expect_{\ppp} \left[ \prod_{u \in \ppp} \left( \frac{\txp}{1 + \thrL{n} \loss{u}{\rx{n}}} + (1 - \txp) \right) \cdot \right. \notag \\
    &\quad \left. \left( 1 - \prod_{u \in \ppp} \left( \frac{\txp}{1 + \thrL{n} \loss{u}{\rx{n}}} + (1 - \txp) \right) \right)^{\tL{n}-1} \right] \notag \\
    &= \sum_{i=0}^{\tL{n}-1} (-1)^i \binom{\tL{n} \! - \! 1}{i}\expect_{\ppp} \left[ \prod_{u \in \ppp} \! \left( \frac{\txp}{1 + \thrL{n} \loss{u}{\rx{n}}} \! + \! (1 - \txp) \! \right) \cdot \right. \notag \\
    &\quad \left. \left( \frac{\txp}{1 + \thrL{n} \loss{u}{\rx{n}}} + (1 - \txp) \right) ^ {\tL{n}-1-i} \right].
    \label{eqn:succ-link}
\end{align}

We finally calculate $\prob \left[ \TL{1} = \tL{1}, \ldots, \TL{N} = \tL{N} \right]$ as shown in~\eqref{eqn:pmf-part} at the top of this page, where $(a)$ follows from applying the probability generating functional of $\ppp$~(see~\cite{haenggi13:book}, (4.8)).
By substituting~\eqref{eqn:pmf-part} into~\eqref{eqn:pmf} we can obtain the PMF of $\T$. However, we can only calculate it for small values of $t$ and $N$ because of computational complexity, since ${t - 1 \choose N - 1}$ values calculated according to~\eqref{eqn:pmf-part} are summed in~\eqref{eqn:pmf}, and each of those terms requires numerically evaluating $\prod_{n=1}^N \tL{n}$ integrals.

Since it is computationally infeasible to obtain the PMF of $\T$ for large values of $t$ and $N$, in the rest of the paper we characterize the relay chain by calculating the expected value and the variance of $\T$ using a computationally tractable method.

\section{Expected Packet Travel Time}
\label{sec:av-time}
In this section, we obtain a closed form expression for the expected value of $\T$. We note that
\begin{equation}
    \expect [ \T ] = \expect \left[ \sum_{n=1}^{N} \TL{n} \right] = \sum_{n=1}^{N} \expect \left[ \TL{n} \right],
    \label{eqn:av-time}
\end{equation}
hence $\expect [\T]$ is completely characterized once having found the expected number of time slots that it takes to travel a single link. This expression is valid for both dependent and independent interference.
\begin{figure*}
    \setcounter{mytempeqncnt}{\value{equation}}
    \setcounter{equation}{10}
    \begin{align}
        &\expect \left[ \TL{m} \TL{n} \right] \overset{m \neq n}{=} \expect_\ppp \left[ \expect\left[ \TL{m} \TL{n} | \ppp \right] \right] = \expect_\ppp \left[ \expect \left[ \TL{m} | \ppp \right] \expect \left[ \TL{n} | \ppp \right] \right] = \expect_\ppp \left[ \frac{1}{\displaystyle \prob \left[ \EvSucc{m} | \ppp \right]} \frac{1}{\displaystyle \prob \left[ \EvSucc{n} | \ppp \right]} \right] \notag \\
        &= \expect_\ppp \left[ \prod_{u \in \ppp} \left(\displaystyle\frac{\txp}{1 + \thrL{m} \loss{u}{\rx{m}}} + (1 - \txp)\right)^{-1} \left(\displaystyle\frac{\txp}{1 + \thrL{n} \loss{u}{\rx{n}}} + (1 - \txp)\right)^{-1} \right] \notag \\
        &= \exp \left( -\dens \int_{\mathbb{R}^2} \left[ 1 - \left(\displaystyle\frac{ \txp}{1 + \thrL{m} \loss{x}{\rx{m}}} + (1 - \txp)\right)^{-1} \left(\displaystyle\frac{ \txp}{1 + \thrL{n} \loss{x}{\rx{n}}} + (1 - \txp)\right)^{-1} \right] \dd x \right).
        \label{eqn:cov-time-dep}
    \end{align}
    \setcounter{equation}{\value{mytempeqncnt}}
    \hrule
\end{figure*}

We remember from the previous section that, in the case of dependent interference, the success events of single transmissions are independent once given a realization of $\ppp$. Hence, for a given realization of $\ppp$, the random variable $\TL{n} | \ppp$ is geometrically distributed with parameter $\prob \left[ \EvSucc{n} | \ppp \right]$. The parameter $\prob \left[ \EvSucc{n} | \ppp \right]$ represents the probability of having a successful transmission on link $\link{n}$ at any time slot conditioned on interferers' positions. Recalling that the expected value of the geometric distribution with parameter $\psi$ is $\psi^{-1}$, we get
\addtocounter{equation}{1}
\begin{align}
    \expect [\TL{n}] &= \expect_{\ppp} [\expect[\TL{n} | \ppp]] = \expect_{\ppp} \left[ \frac{1}{\displaystyle \prob \left[ \EvSucc{n} | \ppp \right]} \right],
    \label{eqn:av-time-dep}
\end{align}
where $\prob \left[ \EvSucc{n} | \ppp \right]$ is obtained in~\eqref{eqn:succ-prob-slot} for the special case of the $i$-th transmission.

By combining~\eqref{eqn:av-time-dep} and~\eqref{eqn:succ-prob-slot} and by applying the probability generating functional of $\ppp$~(see~\cite{haenggi13:book}, (4.8)), we finally get
\begin{align}
    &\expect [\TL{n}] = \expect_{\ppp} \left[ \prod_{u \in \ppp} \frac{1}{\displaystyle\frac{\txp}{1 + \thrL{n} \loss{u}{\rx{n}}} + (1 - \txp)} \right] \notag \\
    &= \exp \left( -\dens \int_{\mathbb{R}^2} \left[ 1 - \frac{1}{\displaystyle\frac{ \txp}{1 + \thrL{n} \loss{x}{\rx{n}}} + (1 - \txp)} \right] \dd x \right).
    \label{eqn:av-time-dep-expl}
\end{align}

If we have independent interference, the single retransmissions on $\link{n}$ are independent, since the positions of interferers at two time slots are independent. Hence, $\TL{n}$ is geometrically distributed with parameter $\prob \left[ \EvSucc{n} \right]$, leading to
\begin{align}
    &\expect [\TL{n}] = \frac{1}{\prob \left[ \EvSucc{n} \right]} = \frac{1}{\expect_{\ppp} \left[ \prob \left[ \EvSucc{n} | \ppp \right] \right]} \notag \\
    &= \frac{1}{\displaystyle \expect_{\ppp} \left[ \prod_{u \in \ppp} \left( \frac{\txp}{1 + \thrL{n} \loss{u}{\rx{n}}} + (1 - \txp) \right) \right]} \notag \\
    &\overset{(a)}{=} \exp {\left( \dens \int_{\mathbb{R}^2} \left[ 1 - \left( \frac{\txp}{1 + \thrL{n} \loss{x}{\rx{n}}} + (1 - \txp) \right) \right] \dd x \right)},
    \label{eqn:av-time-ind-expl}
\end{align}
where, again, $(a)$ follows from applying the probability generating functional of $\ppp$~(see~\cite{haenggi13:book}, (4.8)).

\section{Variance of Packet Travel Time}
\label{sec:var-time}
We now calculate the variance of $\T$. We have
\begin{equation}
    \var[\T] = \expect \left[ \T^2 \right] - \left( \expect \left[ \T \right] \right)^2.
    \label{eqn:var-time-general}
\end{equation}
The value of $\expect \left[ \T \right]$ was obtained in the previous section in~\eqref{eqn:av-time}, with $\expect[\TL{n}]$ given by either~\eqref{eqn:av-time-dep-expl} or~\eqref{eqn:av-time-ind-expl}, and we can calculate $\expect \left[ \T^2 \right]$ as
\begin{align}
    \expect[ T^2] &= \expect \left[ \left( \sum_{n=1}^{N} \TL{n} \right)^2 \right] = \expect \left[ \sum_{m=1}^{N} \TL{m} \sum_{n=1}^{N} \TL{n} \right] \notag \\
    &= \sum_{m=1}^{N} \sum_{n=1}^{N} \expect \left[ \TL{m} \TL{n} \right].
    \label{eqn:av-time-sqr}
\end{align}

For obtaining $\expect[ \TL{m} \TL{n}]$ we distinguish two cases, namely the case $m \neq n$ and the case $m = n$. Consider at first the case where $m \neq n$. We noted in Section~\ref{sec:pmf} that for a given realization of the PPP the success events of two transmissions are independent, which implies that $\TL{m} | \ppp$ and $\TL{n} | \ppp$ are independent. Hence, we have the expressions for $\expect \left[ \TL{m} \TL{n} \right]$ shown in~\eqref{eqn:cov-time-dep} at the top of this page.

If we have independent interference, the single retransmissions are independent, and so are $\TL{m}$ and $\TL{n}$. Hence, we obtain
\begin{equation}
    \expect \left[ \TL{m} \TL{n} \right] = \expect \left[ \TL{m} \right] \expect \left[ \TL{n} \right],
    \label{eqn:cov-time-ind}
\end{equation}
where, again, $\expect \left[ \TL{m} \right]$ and $\expect \left[ \TL{n} \right]$ are obtained according to~\eqref{eqn:av-time-ind-expl}.

Consider now the case where $m = n$. We showed in the previous section that the variable $\TL{n}  | \ppp$ is geometrically distributed with parameter $\prob \left[ \EvSucc{n} \right]$, once having conditioned on a given PPP realization. Recalling that, for a geometrically distributed random variable $\Psi$  with parameter $\psi$, $\expect [\Psi^2] = \frac{2}{\psi^2} - \frac{1}{\psi}$, we obtain the value of $\expect \left[\TL{n}^2 \right]$ shown in~\eqref{eqn:av-time-sqr-dep} at the top of the next page.
\begin{figure*}
    \setcounter{mytempeqncnt}{\value{equation}}
    \setcounter{equation}{18}
    \begin{align}
        &\expect \left[\TL{n}^2 \right] = \expect_{\ppp} [\expect[\TL{n}^2 | \ppp]] = \expect_{\ppp} \left[ \frac{2}{\displaystyle \prob \left[ \EvSucc{n}^2 | \ppp \right]} - \frac{1}{\displaystyle \prob \left[ \EvSucc{n} | \ppp \right]} \right] \notag \\
        &= 2 \expect_{\ppp} \left[ \prod_{u \in \ppp} \frac{1}{\displaystyle\left(\frac{\txp}{ 1 + \thrL{n} \loss{u}{\rx{n}}} + (1 - \txp) \right)^2} \right] - \expect_\ppp \left[ \prod_{u \in \ppp} \frac{1}{\displaystyle\frac{\txp}{1 + \thrL{n} \loss{u}{\rx{n}}} + (1 - \txp)} \right] \notag \\
        &= 2 \exp \left( -\dens \int_{\mathbb{R}^2} \left[ 1 - \frac{1}{\displaystyle\left(\frac{\txp}{ 1 + \thrL{n} \loss{x}{\rx{n}}} + (1 - \txp) \right)^2} \right] \dd x \right) - \exp \left( -\dens \int_{\mathbb{R}^2} \left[ 1 - \frac{1}{\displaystyle\frac{ \txp}{1 + \thrL{n} \loss{x}{\rx{n}}} + (1 - \txp)} \right] \dd x \right).
        \label{eqn:av-time-sqr-dep}
    \end{align}
    \setcounter{equation}{\value{mytempeqncnt}}
    \hrule
\end{figure*}

If we have independent interference, $\TL{n}$ is geometrically distributed with parameter $\prob \left[ \EvSucc{n} \right]$. Hence, we obtain
\begin{align}
    &\expect [\TL{n}^2] = \frac{2}{\displaystyle \prob \left[ \EvSucc{n} \right]^2} - \frac{1}{\displaystyle \prob \left[ \EvSucc{n} \right]} \notag \\
    &= \frac{2}{\displaystyle \left(\expect_\ppp \left[ \prob \left[ \EvSucc{n} | \ppp \right] \right] \right)^2} - \frac{1}{\displaystyle \expect_\ppp \left[ \prob \left[ \EvSucc{n} | \ppp \right] \right]} \notag \\
    &= 2 \exp {\left( 2 \dens \int_{\mathbb{R}^2} \left( 1 - \frac{\txp}{1 + \thrL{n} \loss{x}{\rx{n}}} - (1 - \txp) \right) \dd x \right)} - \notag \\
    &\quad \exp {\left( \dens \int_{\mathbb{R}^2} \left( 1 - \frac{\txp}{1 + \thrL{n} \loss{x}{\rx{n}}} - (1 - \txp) \right) \dd x \right)}.
    \label{eqn:av-time-sqr-ind}
\end{align}

\section{Comparing the Packet Travel Times with Dependent and Independent Interference}
\label{sec:av-var-results}
This section compares $\expect [\T]$ and $\var [\T]$ for the dependent and independent interference cases.
\begin{figure}[t]
    \centering
    \begin{tikzpicture}
    \begin{axis}[
        xlabel={$N$},
        ylabel={$\expect [T]$},
        xmin=1,xmax=20, ymin=0,ymax=50,
        xtick={0,5,...,20},
        xticklabels={0,5,...,20},
        legend style={at={(1,0)}, anchor=south east, font=\footnotesize},
        legend cell align=left,
        legend columns=1,
        font=\footnotesize
     ]
     \addplot[mark=o, mark options={solid, scale=.6}, color=black, dotted, thick] table[x expr=\thisrowno{0}-1, y index =1] {Results/chain-dep-th-0dot1-dist-1-lam-0dot25.dat};
        \label{plt-av-t:dep-lam-0dot25-dist-1}
     \addplot[mark=o, mark options={solid, scale=.6}, color=black, thick] table[x expr=\thisrowno{0}-1, y index =1] {Results/chain-ind-th-0dot1-dist-1-lam-0dot25.dat};
        \label{plt-av-t:ind-lam-0dot25-dist-1}
     \addplot[mark=+, mark options={solid, scale=.8}, color=black, dotted, thick] table[x expr=\thisrowno{0}-1, y index =1] {Results/chain-dep-th-0dot1-dist-1-lam-0dot75.dat};
        \label{plt-av-t:dep-lam-0dot75-dist-1}
     \addplot[mark=+, mark options={solid, scale=.8}, color=black, thick] table[x expr=\thisrowno{0}-1, y index =1] {Results/chain-ind-th-0dot1-dist-1-lam-0dot75.dat};
        \label{plt-av-t:ind-lam-0dot75-dist-1}
     \addplot[mark=square, mark options={solid, scale=.6}, color=black, dotted, thick] table[x expr=\thisrowno{0}-1, y index =1] {Results/chain-dep-th-0dot1-dist-1-lam-1.dat};
        \label{plt-av-t:dep-lam-1-dist-1}
     \addplot[mark=square, mark options={solid, scale=.6}, color=black, thick] table[x expr=\thisrowno{0}-1, y index =1] {Results/chain-ind-th-0dot1-dist-1-lam-1.dat};
        \label{plt-av-t:ind-lam-1-dist-1}
      \addplot[mark=star, mark options={solid, scale=.8}, color=black, dotted, thick] table[x expr=\thisrowno{0}-1, y index =1] {Results/chain-dep-th-0dot1-dist-1-lam-2.dat};
         \label{plt-av-t:dep-lam-2-dist-1}
      \addplot[mark=star, mark options={solid, scale=.8}, color=black, thick] table[x expr=\thisrowno{0}-1, y index =1] {Results/chain-ind-th-0dot1-dist-1-lam-2.dat};
         \label{plt-av-t:ind-lam-2-dist-1}
     \node at (axis cs: 18,26) {$\dens$\,=\,0.25};
     \node at (axis cs: 18,33.5) {$\dens$\,=\,0.75};
     \node at (axis cs: 18,43) {$\dens$\,=\,1};
     \node at (axis cs: 7,40) {$\dens$\,=\,2};
    \end{axis}
\end{tikzpicture}
    \caption{$\expect [\T]$ for different intensities of $\ppp$ for dependent ($\dens = 0.25:$~\ref{plt-av-t:dep-lam-0dot25-dist-1}, $\dens = 0.75:$~\ref{plt-av-t:dep-lam-0dot75-dist-1}, $\dens = 1:$~\ref{plt-av-t:dep-lam-1-dist-1}, $\dens = 2:$~\ref{plt-av-t:dep-lam-2-dist-1}) and independent ($\dens = 0.25:$~\ref{plt-av-t:ind-lam-0dot25-dist-1}, $\dens = 0.75:$~\ref{plt-av-t:ind-lam-0dot75-dist-1}, $\dens = 1:$~\ref{plt-av-t:ind-lam-1-dist-1}, $\dens = 2:$~\ref{plt-av-t:ind-lam-2-dist-1}) interference. Parameters are $\alpha = 3$, $\thr = 0.1$ and $\hop = 1 / N$.}
    \label{fig:av-fix-dist}
\end{figure}

At first, note that the singular path loss of~\cite{haenggi09:outage}
\addtocounter{equation}{1}
\begin{equation}
    \loss{\tx{n}}{\rx{n}} = \| \tx{n} - \rx{n} \| ^ {-\alpha},
    \label{eqn:ploss-idiot}
\end{equation}
produces two drastically different behaviors in the independent and dependent interference cases, when the value $\txp = 1$, i.e., when interferers transmit in every slot. In particular, we find that, in the case of dependent interference, $\expect [\T]$ and $\var [\T]$ are infinite, while they have finite values for the independent interference case. This can be seen by looking at $\expect [\TL{n}]$ for the dependent interference case. By substituting $\txp = 1$ and~\eqref{eqn:ploss-idiot} into~\eqref{eqn:av-time-dep-expl}, we have
\begin{align}
    &\expect [\TL{n}] = \exp \left( \dens \int_{\mathbb{R}^2} \thrL{n} \| x - \rx{n} \| ^ {-\alpha} \dd x \right) \notag \\
    &= \exp \left( 2 \pi \dens \int_{0}^{+\infty} \thrL{n} \rho ^ {-\alpha + 1} \dd \rho \right),
    \label{eqn:av-time-dep-inf}
\end{align}
which is infinite for all values of $\alpha$. By contrast, following the same procedure for the independent interference case by substituting $\txp = 1$ and~\eqref{eqn:ploss-idiot} into~\eqref{eqn:av-time-ind-expl} we find a finite value for $\expect [\TL{n}]$ and $\var [\TL{n}]$.

This behavior is an artifact of the path loss assumption in~\eqref{eqn:ploss-idiot}. For obtaining meaningful results, we consider the non-singular path loss model
\begin{equation}
    \loss{\tx{n}}{\rx{n}} = \min \left(1,  \| \tx{n} - \rx{n} \| ^ {-\alpha}\right).
    \label{eqn:ploss-sense}
\end{equation}

The value of $\expect [\T]$ for a scenario where $\source$ is located at $(0,0)$, $\destination$ at $(1,0)$ and the $N$ relays are equally spaced on the $[0,1]$ segment is shown in Fig.~\ref{fig:av-fix-dist}, for an increasing value of $N$ and different values of $\dens$.

From Fig.~\ref{fig:av-fix-dist} we can see that dependent interference increases the average number of slots required to successfully deliver the packet to $\destination$. The difference, albeit modest, is more significant when $\dens$ increases, i.e., when the density of interferers increases. Fig.~\ref{fig:var-fix-dist} shows $\var [\T]$. It can be seen again that the value for dependent interference is significantly higher than the one for independent interference.
\begin{figure}[t]
    \centering
    \begin{tikzpicture}
    \begin{axis}[
        xlabel={$N$},
        ylabel={$\var [T]$},
        xmin=1,xmax=20, ymin=0,ymax=100,
        xtick={0,5,...,20},
        xticklabels={0,5,...,20},
        legend style={at={(1,0)}, anchor=south east, font=\footnotesize},
        legend cell align=left,
        legend columns=1,
        font=\footnotesize
     ]
     \addplot[mark=o, mark options={solid, scale=.6}, color=black, dotted, thick] table[x expr=\thisrowno{0}-1, y index =2] {Results/chain-dep-th-0dot1-dist-1-lam-0dot25.dat};
        \label{plt-var-t:dep-lam-0dot25-dist-1}
     \addplot[mark=o, mark options={solid, scale=.6}, color=black, thick] table[x expr=\thisrowno{0}-1, y index =2] {Results/chain-ind-th-0dot1-dist-1-lam-0dot25.dat};
        \label{plt-var-t:ind-lam-0dot25-dist-1}
     \addplot[mark=+, mark options={solid, scale=.8}, color=black, dotted, thick] table[x expr=\thisrowno{0}-1, y index =2] {Results/chain-dep-th-0dot1-dist-1-lam-0dot75.dat};
        \label{plt-var-t:dep-lam-0dot75-dist-1}
     \addplot[mark=+, mark options={solid, scale=.8}, color=black, thick] table[x expr=\thisrowno{0}-1, y index =2] {Results/chain-ind-th-0dot1-dist-1-lam-0dot75.dat};
        \label{plt-var-t:ind-lam-0dot75-dist-1}
     \addplot[mark=square, mark options={solid, scale=.6}, color=black, dotted, thick] table[x expr=\thisrowno{0}-1, y index =2] {Results/chain-dep-th-0dot1-dist-1-lam-1.dat};
        \label{plt-var-t:dep-lam-1-dist-1}
     \addplot[mark=square, mark options={solid, scale=.6}, color=black, thick] table[x expr=\thisrowno{0}-1, y index =2] {Results/chain-ind-th-0dot1-dist-1-lam-1.dat};
        \label{plt-var-t:ind-lam-1-dist-1}
     \node at (axis cs: 15.5,41) {dep., $\dens$\,=\,0.25};
     \node at (axis cs: 13,75) {dep., $\dens$\,=\,0.75};
     \node at (axis cs: 7.5,80) {dep., $\dens$\,=\,1};
     \node at (axis cs: 16,8) {ind., $\dens$\,=\,0.25};
     \node at (axis cs: 17.8,28) {ind., $\dens$\,=\,0.75};
     \node at (axis cs: 17,53) {ind., $\dens$\,=\,1};
    \end{axis}
\end{tikzpicture}
    \caption{$\var [\T]$ for different intensities of $\ppp$ for dependent ($\dens = 0.25:$~\ref{plt-var-t:dep-lam-0dot25-dist-1}, $\dens = 0.75:$~\ref{plt-var-t:dep-lam-0dot75-dist-1}, $\dens = 1:$~\ref{plt-var-t:dep-lam-1-dist-1}) and independent ($\dens = 0.25:$~\ref{plt-var-t:ind-lam-0dot25-dist-1}, $\dens = 0.75:$~\ref{plt-var-t:ind-lam-0dot75-dist-1}, $\dens = 1:$~\ref{plt-var-t:ind-lam-1-dist-1}) interference. Parameters are $\alpha = 3$, $\thr = 0.1$ and $\hop = 1 / N$.}
    \label{fig:var-fix-dist}
\end{figure}
\begin{figure}[t]
    \centering
    \begin{tikzpicture}
   \begin{axis}[
       xlabel={$N$},
       ylabel={$\var [T]$},
       xmin=0,xmax=10, ymin=1,ymax=1000,
       xtick={0,2,...,10},
       xticklabels={0,2,...,10},
       legend style={at={(1,0)}, anchor=south east, font=\footnotesize},
       legend cell align=left,
       legend columns=1,
       font=\footnotesize
   ]

   \addplot[mark=o, mark options={solid, scale=.8}, color=black, dashed, thick] table[x expr=\thisrowno{0}-1, y index =2] {Results/chain-dep-th-0dot1-hop-0dot1-lam-2.dat};
      \label{plt-av-t:dep-lam-2-hop-0dot1}
   \addplot[mark=o, mark options={solid, scale=.8}, color=black, thick] table[x expr=\thisrowno{0}-1, y index =2] {Results/chain-ind-th-0dot1-hop-0dot1-lam-2.dat};
      \label{plt-av-t:ind-lam-2-hop-0dot1}
   \addplot[mark=+, mark options={solid, scale=1}, color=black, dashed, thick] table[x expr=\thisrowno{0}-1, y index =2] {Results/chain-dep-th-0dot1-hop-0dot25-lam-2.dat};
       \label{plt-av-t:dep-lam-2-hop-0dot25}
   \addplot[mark=square, mark options={solid, scale=.8}, color=black, dashed, thick] table[x expr=\thisrowno{0}-1, y index =2] {Results/chain-dep-th-0dot1-hop-0dot5-lam-2.dat};
      \label{plt-av-t:dep-lam-2-hop-0dot5}
   \addplot[mark=star, mark options={solid, scale=1}, color=black, dashed, thick] table[x expr=\thisrowno{0}-1, y index =2] {Results/chain-dep-th-0dot1-hop-0dot75-lam-2.dat};
      \label{plt-av-t:dep-lam-2-hop-0dot75}
   \addplot[mark=diamond, mark options={solid, scale=1}, color=black, dashed, thick] table[x expr=\thisrowno{0}-1, y index =2] {Results/chain-dep-th-0dot1-hop-1-lam-2.dat};
      \label{plt-av-t:dep-lam-2-hop-1}
      \draw[thick, ->] (axis cs: 6.5,600) node[anchor=north east] {dep, $L = 0.1, 0.25, 0.5, 0.75, 1$} -- (axis cs: 6.5,300) ;
      \node at (axis cs: 6,100) {ind., $L = 0.1, 0.25, 0.5, 0.75, 1$};
   \end{axis}
\end{tikzpicture}
    \caption{$\var [\T]$ for different link length values for dependent ($\hop = 0.1:$~\ref{plt-av-t:dep-lam-2-hop-0dot1}, $\hop = 0.25:$~\ref{plt-av-t:dep-lam-2-hop-0dot25}, $\hop = 0.5:$~\ref{plt-av-t:dep-lam-2-hop-0dot5}, $\hop = 0.75:$~\ref{plt-av-t:dep-lam-2-hop-0dot75}, $\hop = 1:$~\ref{plt-av-t:dep-lam-2-hop-1}) and independent ($\hop = 0.1, 0.25, 0.5, 0.75, 1:$~\ref{plt-av-t:ind-lam-2-hop-0dot1}) interference. Parameters are $\alpha = 3$, $\thr = 0.1$ and $\dens = 2$. Note that $\var[\T]$ does not depend on $L$ in the independent interference case for $L \leq 1$, since we have $\expect [\TL{m} \TL{n}] = \expect [\TL{m}] \expect [\TL{n}]$, and $\thrL{m} = 1$ for $L \leq 1$. Conversely, when interference is dependent the quantities $\rx{n}$ and $\rx{m}$ appear in the same integral for calculating $\expect [\TL{m} \TL{n}]$, hence $\var[\T]$ depends on $\| \rx{n} - \rx{m}\| = L$.}
    \label{fig:var-fix-hop}
    \vspace{-0.2cm}
\end{figure}

Results for $\var[T]$ for dependent and independent interference when all links have a length of $\hop$, and hence the total distance between $\source$ and $\destination$ is $N \hop$, are shown in Fig.~\ref{fig:var-fix-hop}.

From Figs.~\ref{fig:var-fix-dist} and~\ref{fig:var-fix-hop}, one can notice that another difference between the dependent and independent interference cases is the relation between $\var [\T]$ and $N$. In particular, we observe that in the dependent interference case $\var [\T]$ grows faster than linearly with $N$, while in the independent interference case it grows linearly with $N$, as in this case $\var[\T] = \sum_{n=1}^{N} \var[\TL{n}]$. However, one can observe that the growth of $\var [\T]$ with $N$ becomes almost linear when the number of relays increases while maintaining the link length constant. This is because the covariance between the interference powers at two links $\link{m}$ and $\link{n}$ decreases when the distance between the links increases, as shown in Fig.~\ref{fig:cov-vs-dist}. Hence, two distant links are almost independent, and $\expect [\TL{m} \TL{m}]$ is very close to $\expect [\TL{m}] \expect [\TL{n}]$. These two terms cancel out in the expression of the variance, and thus the only superlinear contribution is given by links that are close, in the sense that their receptions are dependent. However, when having many relays located at a fixed distance, each node is close enough, in this sense, to only a small fraction of the rest of the nodes, and their quadratic contribution is not significant. Hence, the value of $\var [\T]$ is very close to being linear with respect to $N$. This result is discussed in more depth in the next section.
\begin{figure}[t]
    \centering
    \begin{tikzpicture}
    \begin{semilogyaxis}[
        xlabel={$\| \rx{m} - \rx{n} \|$},
        ylabel={$\cov [\TL{m} \TL{n}] = \expect [\TL{m} \TL{n}] - \expect [\TL{m}] \expect [\TL{n}]$},
        xmin=0,xmax=10, ymin=1e-3,ymax=10,
        xtick={0,2,...,10},
        xticklabels={0,2,...,10},
        legend style={at={(1,0)}, anchor=south east, font=\footnotesize},
        legend cell align=left,
        legend columns=1,
        font=\footnotesize
     ]
     \addplot[mark=*, mark options={solid, scale=.6}, mark repeat=5, color=black, dotted, thick] table[x index=0, y expr=\thisrowno{1}-\thisrowno{2}] {Results/cov-th-0dot1-hop-1-lam-0dot5.dat};
        \label{plt-cov:lam-0dot5}
     \addplot[mark=+, mark options={solid, scale=1}, mark repeat=5, color=black, dotted, thick] table[x index=0, y expr=\thisrowno{1}-\thisrowno{2}] {Results/cov-th-0dot1-hop-1-lam-1.dat};
        \label{plt-cov:lam-1}
     \addplot[mark=square*, mark options={solid, scale=.6}, mark repeat=5, color=black, dotted, thick] table[x index=0, y expr=\thisrowno{1}-\thisrowno{2}] {Results/cov-th-0dot1-hop-1-lam-1dot5.dat};
        \label{plt-cov:lam-1dot5}
     \addplot[mark=star, mark options={solid, scale=1}, mark repeat=5, color=black, dotted, thick] table[x index=0, y expr=\thisrowno{1}-\thisrowno{2}] {Results/cov-th-0dot1-hop-1-lam-2.dat};
        \label{plt-cov:lam-2}
     \node[anchor=west] at (axis cs: 7.2,0.06) {$\dens$\,=\,0.5};
     \node[anchor=west] at (axis cs: 7.2,0.016) {$\dens$\,=\,1};
     \node[anchor=west] at (axis cs: 7.2,0.004) {$\dens$\,=\,1.5};
     \node at (axis cs: 6,0.002) {$\dens$\,=\,2};
    \end{semilogyaxis}
\end{tikzpicture}
    \caption{$\cov [\TL{m} \TL{n}] =  \expect [\TL{m} \TL{n}] - \expect [\TL{m}] \expect [\TL{n}]$ for different densities of $\ppp$ ($\dens = 0.5:$~\ref{plt-cov:lam-0dot5}, $\dens = 1:$~\ref{plt-cov:lam-1}, $\dens = 1.5:$~\ref{plt-cov:lam-1dot5}, $\dens = 2:$~\ref{plt-cov:lam-2}). Parameters are $\alpha = 3$, $\thr = 0.1$ and $\hop = 1$.}
    \label{fig:cov-vs-dist}
    \vspace{-0.2cm}
\end{figure}

\section{Speed of a Packet Traveling a Uniform Chain}
\label{sec:speed}
In the previous section we showed that the covariance $\cov \left[ \TL{m}, \TL{n} \right] = \expect [\TL{m} \TL{n}] - \expect [\TL{m}] \expect [\TL{n}]$ decreases rapidly with the distance between two links in a uniform chain scenario, i.e., when all the links have the same length $\hop$. Motivated by Fig.~\ref{fig:cov-vs-dist}, we can approximate the covariance value by assuming that  $\cov \left[ \TL{n},  \TL{n+k} \right] = 0$ if $k > K$, that is, assuming that two links that have at least $K$ other links between them are uncorrelated. We also have $\cov \left[ \TL{n},  \TL{n+k} \right] \leq \cov \left[ \TL{(\cdot)} \TL{(\cdot)+1} \right] = \covbound, \forall n, k$, where we denote the bounding value of the covariance with $\covbound$. We also use $\TL{(\cdot)}$ for representing the number of slots required to travel a single link, since it does not depend on the link index, given that all links have the same length $\hop$. We can then rewrite $\var [\T]$ as
\begin{align}
    \var[\T] &= \sum_{n=1}^{N} \var \left[ \TL{n} \right] + \sum_{m=1}^{N} \sum_{\substack{n=1 \\ n \neq m}}^{N} \cov \left[ \TL{m}\TL{n} \right] \notag \\
    &= \sum_{n=1}^{N} \var \left[ \TL{n} \right] + 2 \sum_{n=1}^{N} \sum_{k=1}^{\min (N-n, K)} \cov \left[ \TL{n}\TL{n+k} \right], \notag \\
    &\leq N \var \left[ \TL{(\cdot)} \right] + 2 N \min (N-n, K) \covbound,
    \label{eqn:var-time-with-cov}
\end{align}
hence, one can see from~\eqref{eqn:var-time-with-cov} that $\var [\T]$ grows linearly with $N$ when distant links are uncorrelated.

We define the random variable $\speed$ representing the average speed of a packet as
\begin{equation}
    \speed = \frac{N \hop}{\T},
    \label{eqn:speed}
\end{equation}
which is the ratio between the distance that the packet travels and the time it takes the packet to reach its destination. One can see that, in order to obtain the PMF of the average speed of the packet, we need to calculate the PMF of $\T$. However, we can simplify the calculation when the number of relays is infinite, and the covariance between two links is zero when they are sufficiently separated, so that~\eqref{eqn:var-time-with-cov} holds.

In particular, let us consider $\speed^{-1} = \frac{\T}{N \hop}$. We have
\begin{align}
    \expect \left[ \speed^{-1} \right] &= \expect \left[ \frac{\T}{N \hop} \right] = \frac{\sum_{n = 1}^{N} \expect [\TL{n}]}{N \hop} = \frac{\expect [\TL{(\cdot)}]}{\hop},
    \label{eqn:av-uniform-chain} \\
    \var \left[ \speed^{-1} \right] &= \var \left[ \frac{\T}{N \hop} \right] = \left(\frac{1}{N \hop}\right)^2 \var [\T].
    \label{eqn:var-uniform-chain}
\end{align}

\begin{figure}[t]
    \centering
    \begin{tikzpicture}
    \begin{axis}[
        xlabel={$\hop$},
        ylabel={$\displaystyle \frac{\hop}{\expect[\TL{(\cdot)}]} $},
        xmin=1,xmax=5, ymin=0,ymax=1.2,
        xtick={1,...,5},
        xticklabels={1,...,5}
     ]
     \addplot[mark=o, mark options={solid, scale=.6}, mark repeat=2, color=black, dotted, thick] table[x index=0, y index =1] {Results/speed-th-0dot2-p-0dot25-lam-0dot25.dat};
        \label{plt-speed:dep-lam-0dot25-p-0dot25}
     \addplot[mark=star, mark options={solid, scale=.8}, mark repeat=2, color=black, dotted, thick] table[x index=0, y index =1] {Results/speed-th-0dot2-p-0dot5-lam-0dot25.dat};
        \label{plt-speed:dep-lam-0dot25-p-0dot5}
     \addplot[mark=square*, mark options={solid, scale=.6}, mark repeat=2, color=black, dotted, thick] table[x index=0, y index =1] {Results/speed-th-0dot2-p-0dot75-lam-0dot25.dat};
        \label{plt-speed:dep-lam-0dot25-p-0dot75}
     \addplot[mark=o, mark options={solid, scale=.6}, mark repeat=2, color=black, thick] table[x index=0, y index =2] {Results/speed-th-0dot2-p-0dot25-lam-0dot25.dat};
        \label{plt-speed:ind-lam-0dot25-p-0dot25}
     \addplot[mark=star, mark options={solid, scale=.8}, mark repeat=2, color=black, thick] table[x index=0, y index =2] {Results/speed-th-0dot2-p-0dot5-lam-0dot25.dat};
        \label{plt-speed:ind-lam-0dot25-p-0dot5}
     \addplot[mark=square*, mark options={solid, scale=.6}, mark repeat=2, color=black, thick] table[x index=0, y index =2] {Results/speed-th-0dot2-p-0dot75-lam-0dot25.dat};
        \label{plt-speed:ind-lam-0dot25-p-0dot75}
     \draw (axis cs: 3,0.65) ellipse [x radius=.3,  y radius=.05];
     \draw (axis cs: 2.65,0.23) ellipse [x radius=.3,  y radius=.05];
     \draw (axis cs: 1.8,0.34) ellipse [x radius=.3,  y radius=.05];
     \node[anchor=west] at (axis cs: 3.3,0.65) {$\txp \! = \! 0.25$};
     \node[anchor=west] at (axis cs: 3.0,0.23) {$\txp \! = \! 0.5$};
     \node[anchor=west] at (axis cs: 0.98,0.25) {$\txp \! = \! 0.75$};
    \end{axis}
\end{tikzpicture}
    \vspace{-0.2cm}
    \caption{Asymptotic average speed of a packet along the relay chain for $N \rightarrow \infty$ and for different transmission probabilities $\txp$ and for dependent ($\txp = 0.25$:~\ref{plt-speed:dep-lam-0dot25-p-0dot25}, $\txp = 0.5$:~\ref{plt-speed:dep-lam-0dot25-p-0dot5}, $\txp = 0.75$:~\ref{plt-speed:dep-lam-0dot25-p-0dot75}) and independent interference ($\txp = 0.25$:~\ref{plt-speed:ind-lam-0dot25-p-0dot25}, $\txp = 0.5$:~\ref{plt-speed:ind-lam-0dot25-p-0dot5}, $\txp = 0.75$:~\ref{plt-speed:ind-lam-0dot25-p-0dot75}). Parameters are $\alpha = 3$, $\thr = 0.2$ and $\dens = 0.25$.}
    \label{fig:speed}
    \vspace{-0.2cm}
\end{figure}
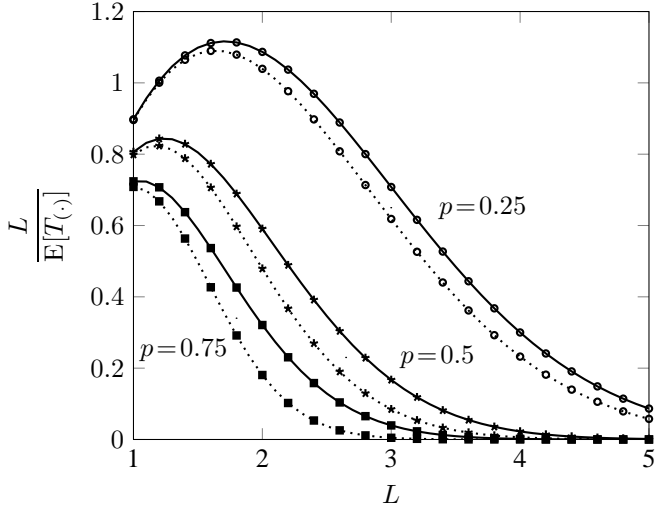

Using Chebyshev's inequality, we have, for some $\epsilon > 0$,
\begin{equation}
    \prob \left[ \left\lvert \speed^{-1} - \expect \left[ \speed^{-1} \right] \right\rvert > \epsilon \right] \leq \frac{\var[\speed^{-1}]}{\epsilon^2}
    \label{eqn:speed-infinite}
\end{equation}
and by substituting~\eqref{eqn:av-uniform-chain} and~\eqref{eqn:var-uniform-chain} into~\eqref{eqn:speed-infinite} we obtain
\begin{equation}
    \prob \left[ \left\lvert \speed^{-1} - \frac{\expect [\TL{(\cdot)}]}{\hop} \right\rvert > \epsilon \right] \leq \frac{\var [\T]}{N^2 \hop^2 \epsilon^2}.
    \label{eqn:speed-infinite-intermediate}
\end{equation}
We now consider the asymptotic scenario where the number of relays is infinite, leading to
\begin{equation}
    \lim_{N \rightarrow \infty} \! \prob \left[ \left\lvert \speed^{-1} - \frac{\expect [\TL{(\cdot)}]}{\hop} \right\rvert > \epsilon \right] \! \leq \! \lim_{N \rightarrow \infty} \frac{\var [\T]}{N^2 \hop^2 \epsilon^2} \overset{(a)}{=} 0,
    \label{eqn:speed-infinite-final}
\end{equation}
where $(a)$ follows from recalling that $\var[\T]$ grows linearly with $N$. Hence, in the asymptotic case, the average speed of a packet converges in probability to the value $\frac{\hop}{\expect [\TL{(\cdot)}]}$ (see~\cite{shao2003matstat}, Theorem 1.10). The value of $\speed$ is shown in Fig.~\ref{fig:speed}, and we observe that in the dependent interference scenario the packet travels the chain with a lower average speed.

\section{Conclusions}
\label{sec:conclusions}
This paper analyzed the time it takes a packet to traverse a chain of relays with surrounding interferers. Derivations assume Rayleigh fading, Poisson distributed interferers, slotted ALOHA, and infinite packet retransmissions for each hop. The spatiotemporal dependence of interference has significant impact on the travel time. In particular, if the interferers' positions remain fixed for the entire duration, a packet is on average slower as if they are positioned anew in each slot. Also the variance of the packet travel time is higher in the first case. These facts should be taken into consideration when analyzing packet propagation in wireless multi-hop networks.

\section*{Acknowledgments}
This work has been supported by the Austrian Science Fund (FWF) grant P24480-N15, KWF/EFRE grants 20214/15935/23108 (RELAY) and 20214/20777/31602 (Research Days), the European social fund, and Greek national funds through the program ``Education and Lifelong Learning'' of the NSRF program ``THALES investing in knowledge society through the European Social Fund''.

\bibliographystyle{ieeetr}    
\bibliography{bettstetter,paper,alessandro}

\begin{thebibliography}{10}

\bibitem{xie2004relay}
L.-L. Xie and P.~Kumar, ``An achievable rate for the multiple level relay
  channel,'' in {\em Proc.~IEEE Intern. Symp. on Inform. Theory ({ISIT})}, June
  2004.

\bibitem{farhadi2009multihop}
G.~Farhadi and N.~Beaulieu, ``On the ergodic capacity of multi-hop wireless
  relaying systems,'' {\em {IEEE} Trans. Wireless Commun.}, vol.~8,
  pp.~2286--2291, May 2009.

\bibitem{boyer2004mutlihop}
J.~Boyer, D.~Falconer, and H.~Yanikomeroglu, ``Multihop diversity in wireless
  relaying channels,'' {\em {IEEE} Trans. Commun.}, vol.~52, pp.~1820--1830,
  Oct. 2004.

\bibitem{haenggi09:interference}
M.~Haenggi and R.~Ganti, {\em Interference in Large Wireless Networks}.
\newblock Now Publishers, 2009.

\bibitem{haenggi09:outage}
M.~Haenggi, ``Outage, local throughput, and capacity of random wireless
  networks,'' {\em {IEEE} Trans. Commun.}, vol.~8, pp.~4350--4359, Aug. 2009.

\bibitem{ganti09:interf-correl}
R.~Ganti and M.~Haenggi, ``Spatial and temporal correlation of the interference
  in {ALOHA} ad hoc networks,'' {\em {IEEE} Commun. Lett.}, vol.~13,
  pp.~631--633, Sept. 2009.

\bibitem{schilcher12:interfcor}
U.~Schilcher, C.~Bettstetter, and G.~Brandner, ``Temporal correlation of
  interference in wireless networks with \mbox{Rayleigh} fading,'' {\em IEEE
  Trans. Mobile Comput.}, vol.~11, pp.~2109--2120, Dec. 2012.

\bibitem{tanbourgi2013coop}
R.~Tanbourgi, H.~J{\"a}kel, and F.~Jondral, ``Cooperative relaying in a poisson
  field of interferers: A diversity order analysis,'' in {\em Proc.~IEEE
  Intern. Symp. on Inform. Theory ({ISIT})}, July 2013.

\bibitem{schilcher2013coop}
U.~Schilcher, S.~Toumpis, A.~Crismani, G.~Brandner, and C.~Bettstetter, ``How
  does interference dynamics influence packet delivery in cooperative
  relaying?,'' in {\em ACM International Conference on Modeling, Analysis and
  Simulation of Wireless and Mobile Systems (MSWiM)}, Nov. 2013.

\bibitem{schilcher2013temporal}
U.~Schilcher, G.~Brandner, and C.~Bettstetter, ``Temporal correlation of
  interference: Cases with correlated traffic,'' {\em Proc. Intern. ITG Conf.
  Syst., Commun. and Coding (SCC)}, 2013.

\bibitem{haenggi2014divpol}
M.~Haenggi and R.~Smarandache, ``{Diversity polynomials for the analysis of
  temporal correlations in wireless networks},'' {\em {IEEE} Trans. Wireless
  Commun.}, vol.~12, pp.~5940--5951, Nov. 2013.

\bibitem{bacelli2006aloha}
F.~Baccelli, B.~Blaszczyszyn, and P.~Muhlethaler, ``An {Aloha} protocol for
  multihop mobile wireless networks,'' {\em {IEEE} Trans. Inform. Theory},
  vol.~52, pp.~421--436, Feb. 2006.

\bibitem{haenggi2013local}
M.~Haenggi, ``The local delay in poisson networks,'' {\em {IEEE} Trans. Inform.
  Theory}, vol.~59, pp.~1788--1802, Mar. 2013.

\bibitem{haenggi2013moblocal}
Z.~Gong and M.~Haenggi, ``The local delay in mobile poisson networks,'' {\em
  {IEEE} Trans. Wireless Commun.}, vol.~12, no.~9, pp.~4766--4777, 2013.

\bibitem{haenggi2013delay}
K.~Stamatiou and M.~Haenggi, ``{Delay characterization of multihop transmission
  in a Poisson field of interference},'' {\em {IEEE/ACM} Trans. Networking}.
\newblock Accepted. Available at
  \url{http://www.nd.edu/~mhaenggi/pubs/ton14.pdf}.

\bibitem{stoyan95}
D.~Stoyan, W.~S. Kendall, and J.~Mecke, {\em Stochastic Geometry and Its
  Applications}, ch.~2.4.
\newblock John Wiley \& Sons Ltd, 1995.

\bibitem{haenggi13:book}
M.~Haenggi, {\em Stochastic Geometry for Wireless Networks}.
\newblock Cambridge University Press, 2013.

\bibitem{shao2003matstat}
J.~Shao, {\em Matematical Statistics}.
\newblock Springer, 2003.

\end{thebibliography}

\end{document}